%% file: template.tex
\documentclass{article}

\usepackage{arxiv}

\usepackage[utf8]{inputenc} 
\usepackage[T1]{fontenc}    
\usepackage{hyperref}       
\usepackage{url}            
\usepackage{booktabs}       
\usepackage{amsfonts}       
\usepackage{nicefrac}       
\usepackage{microtype}      
\usepackage{lipsum}
\usepackage{graphicx}
\graphicspath{ {./images/} }

\usepackage[table,dvipsnames]{xcolor}
\usepackage{pifont}

\usepackage[square,numbers]{natbib}
\newcommand*\rot{\rotatebox{90}}
\newcommand*\OK{\ding{51}}
\newcommand*\NO{\ding{55}}

\usepackage{tikz}
\usepackage{tcolorbox}
\definecolor{cell}{HTML}{B0BEC5}

\definecolor{measured}{HTML}{6fa8dc}
\definecolor{condition}{HTML}{93c47d}
\definecolor{fixed}{HTML}{ffd966}

\definecolor{measured}{HTML}{66c2a5}
\definecolor{condition}{HTML}{5e4fa2}
\definecolor{fixed}{HTML}{f46d43}

\definecolor{user}{HTML}{DF7676}
\definecolor{explanation}{HTML}{6999B5}
\definecolor{model}{HTML}{E99D00}
\definecolor{te}{HTML}{78a34e}

\definecolor{C0}{HTML}{F5F5F5}
\definecolor{C1}{HTML}{E8E8E8}
\definecolor{C2}{HTML}{DCDCDC}
\definecolor{C3}{HTML}{C0C0C0}
\definecolor{C4}{HTML}{A0A0A0}
\definecolor{C5}{HTML}{808080}
\definecolor{C6}{HTML}{606060}

\definecolor{CA}{HTML}{F5F5F5}
\definecolor{CU}{HTML}{E8E8E8}
\definecolor{CS}{HTML}{DCDCDC}
\definecolor{CD}{HTML}{C0C0C0}
\definecolor{CR}{HTML}{A0A0A0}
\definecolor{CJ}{HTML}{808080}
\definecolor{CC}{HTML}{606060}

\newcommand{\co}{
\begin{tikzpicture}[every node/.style={inner sep=0,outer sep=.0}]
    \fill [rounded corners=0.03cm,fill=condition] (0,0)--(1em,0)--(1em,1em)--(0,1em)--cycle;
\end{tikzpicture}
}

\newcommand{\f}{
  \begin{tikzpicture}[every node/.style={inner sep=0,outer sep=0}]
      \fill [rounded corners=0.03cm,fill=fixed] (0,0)--(1em,0)--(1em,1em)--(0,1em)--cycle;
  \end{tikzpicture}
}

\newcommand{\dff}{
  \begin{tikzpicture}[every node/.style={inner sep=0,outer sep=0}]
    \node (O) [rounded corners=0.03cm,draw=fixed,ultra thick,align=center,anchor=center,minimum width=0.32cm, minimum height=0.32cm] at (0.1,0.1) {};
  \end{tikzpicture}
}

\newcommand{\m}{
    \begin{tikzpicture}[every node/.style={inner sep=0,outer sep=0}]
        \fill [rounded corners=0.03cm,fill=measured] (0,0)--(1em,0)--(1em,1em)--(0,1em)--cycle;
    \end{tikzpicture}
}

\newcommand{\cm}{
    \begin{tikzpicture}[every node/.style={inner sep=0,outer sep=0}]
        \fill [rounded corners=0.03cm,fill=measured] (0,0)--(0,1em)--(1em,1em)--cycle;
        \fill [rounded corners=0.03cm,fill=condition] (0,0)--(1em,0)--(1em,1em)--cycle;
    \end{tikzpicture}
}

\newcommand{\quant}[1]{
    \begin{tikzpicture}[every node/.style={inner sep=0pt,outer sep=0}]
        \node (O) [rounded corners=0.03cm,fill=C#1,align=center,anchor=center,minimum width=0.32cm, minimum height=0.32cm] at (0.1,0.1) {\scriptsize #1};
    \end{tikzpicture}
}

\newcommand{\dfquant}[1]{
    \begin{tikzpicture}[every node/.style={inner sep=0pt,outer sep=0}]
        \node (O) [rounded corners=0.03cm,draw=C#1,ultra thick,align=center,anchor=center,minimum width=0.32cm, minimum height=0.32cm] at (0.1,0.1) {\scriptsize #1};
    \end{tikzpicture}
}

\newcommand{\tablecite}[1]{
\citeauthor{#1} (\citeyear{#1}) \cite{#1}
}

\usepackage{enumitem}
\setlist[description]{leftmargin=0cm,labelindent=0cm}
\newcommand{\bulletitem}[2]{
\item[\textcolor{#2}{$\bullet$}~#1~---]
}

\title{Should We Trust (X)AI? Design Dimensions \\ for Structured Experimental Evaluations}

\author{
 Fabian Sperrle \\
  Univeristy of Konstanz \\
  \texttt{fabian.sperrle@uni.kn} \\
   \And
 Mennatallah El-Assady \\
 University of Konstanz \\
  \texttt{mennatallah.el-assady@uni.kn} \\
  \And
  Grace Guo \\
  Georgia Institute of Technology \\
  \texttt{gguo31@gatech.edu} \\
  \And
  Duen Horng Chau \\
  Georgia Institute of Technology \\
  \texttt{polo@gatech.edu} \\
   \And
  Alex Endert \\
  Georgia Institute of Technology \\
  \texttt{endert@gatech.edu} \\
  \And
  Daniel Keim \\
  University of Konstanz \\
  \texttt{daniel.keim@uni.kn} \\
}

\begin{document}
\maketitle
\begin{abstract}
This paper systematically derives design dimensions for the structured evaluation of explainable artificial intelligence (XAI) approaches. These dimensions enable a descriptive characterization, facilitating comparisons between different study designs. They further structure the design space of XAI, converging towards a precise terminology required for a rigorous study of XAI. Our literature review differentiates between comparative studies and application papers, revealing methodological differences between the fields of machine learning, human-computer interaction, and visual analytics. Generally, each of these disciplines targets specific parts of the XAI process. Bridging the resulting gaps enables a holistic evaluation of XAI in real-world scenarios, as proposed by our conceptual model characterizing bias sources and trust-building. Furthermore, we identify and discuss the potential for future work based on observed research gaps that should lead to better coverage of the proposed model.
\end{abstract}

\section{Introduction}

\begin{figure}[h]
    \centering
    \includegraphics[width=\textwidth]{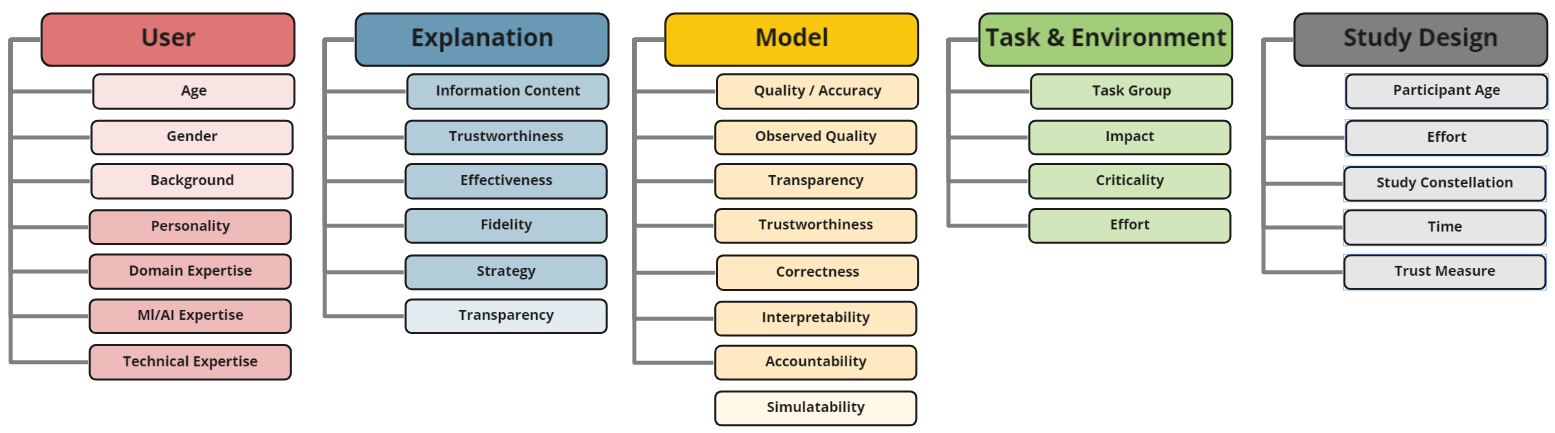}
    \caption{The design dimensions of (X)AI evaluation can be organized in the five groups \emph{user, explanation, model, task \& environment,} and \emph{study design}. The individual dimensions have been collected from related work and are defined in \autoref{sec:dimensions}. }
    \label{fig:teaser}
\end{figure}

With the recent leaps in the performance and utility of Artificial Intelligence (AI) across a variety of applications, the demand for understanding their decision-making rationale is on the rise.   \textit{Explainable Artificial Intelligence} (XAI) is the study of making the decision-making processes of AI models explainable. Explanations not only can help foster trust among novice users but are also valuable tools when \emph{discovering, improving, controlling, or justifying}~\cite{adadi_peeking_2018} the machine learning models powering AI. Consequently, many different approaches to explaining AI have emerged in recent years~\cite{abdul_trends_2018}. XAI encompasses explanations throughout the whole \textit{process} of machine learning from the raw data to presenting the discovered relations and patterns to the user. Within this process, XAI methods focus on explaining the data, the AI model, or presenting the output of the XAI method to the user~\cite{spinner_explAIner_2019}. 

A central task for XAI is calibrating trust in the context of complex machine learning models and processes that are not always intelligible. The fact that it is often difficult for humans to comprehend the inner workings of models raises many questions towards methods claiming to provide explanations: Are they valid? Do they calibrate user trust appropriately, or introduce bias? For which data types and tasks are they applicable, and in which environments?  In focused application contexts, some approaches can evaluate the general propagation of these effects to derive suitable architectures. However, in the general context of evaluating (X)AI and finding answers to these questions requires comparing different ways to generate, design, and present explanations to different user types. While catering explanations to their intended audience concerning, for example, complexity and information density might seem straightforward, it is not always easy due to the complexity of various psychological processes. Additionally, Miller et al. state that ``most of us as AI researchers are building explanatory agents for ourselves, rather than for the intended users''~\cite{miller_explainable_2017} and that XAI is more likely to be successful ``if evaluation of these models is focused more on people than on technology.''~\cite{miller_explainable_2017}

\citet{doshi-velez_towards_2017} identify that interpretability is mostly evaluated in the context of a concrete application, or assumed to be ``given'' thanks to the use of a particular model class, stating that  ``To large extent, both evaluation approaches rely on some notion of ``you'll know it when you see it''.''~\cite{doshi-velez_towards_2017} According to \citet{guidotti_survey_2018}, ``In the state of the art a small set of existing interpretable models is recognized: decision tree, rules, linear models. These models are considered easily understandable and interpretable for humans''~\cite{guidotti_survey_2018}. \citet{poursabzi-sangdeh_manipulating_2018}, however, evaluated such white-box linear models and found that transparency can be overwhelming, possibly due to information overload.

In this paper, we attempt to generate a comprehensive overview of the design dimensions for structured experimental evaluation of XAI methods. To that end, we contribute a literature review of the past five years of (X)AI evaluation research in the human-computer-interaction and visualization communities and report on their considerations for XAI studies and their design. 
In addition, we review a number of application papers that do not feature comparative study designs, but present user evaluations showcasing the effectiveness of XAI methods. 
From this review, we generalize and contribute five groups of design dimensions for future evaluation studies: \emph{personal characteristics, explanation, model, task and environment,} and \emph{study setup}.
From this literature review we derive not only research gaps, but also opportunities and implications for future work. Further, we contribute a dependency model outlining all actors that must be considered when designing experiments evaluating trust in (X)AI. This model also describes the processes of bias propagation and trust building in (X)AI.

\section{Background and Related Work}
\label{sec:rel_work}

Already in 1994, Muir presented ``a theoretical model of human trust in machines.''~\cite{muir_trust_1994}. She collected multiple definitions of trust in humans, showcasing the different aspects of trust and suggesting that ``trust is a multidimensional construct''~\cite{muir_trust_1994} and that calibration and re-calibration of trust are necessary over time. More recently, Robbins surveyed psychological literature on trust, finding it to be fragmented with multidimensional definitions. He instead suggest defining trust in terms of ``actor A's beliefs, actor B's trustworthiness, the matter(s) at hand, and unknown outcomes.''~\cite{robbins_what_2016}

In addition to these theoretical considerations on trust, \citet{hancock_meta-analysis_2011} provide a meta-analysis on factors influencing the trust of humans in robots. This work was later extended to automation in general by \citet{schaefer_meta-analysis_2016}. Neither of the works places emphasis on how these factors can be evaluated. Their dimensions might, however, also be relevant for trust of humans in AI and can be grouped similarly to those presented in \autoref{sec:dimensions}. 

Recent work from computer science provided a conceptual framework for designing XAI~\cite{wang_designing_2019}. The authors reviewed work from psychology and philosophy and suggested how XAI should be designed to avoid cognitive biases. These guidelines are applied and evaluated in an application for clinical decision making. More generally, \citet{doshi-velez_towards_2017} provide a taxonomy for the evaluation of interpretable machine learning systems, identifying three types of studies: application-grounded, human-grounded, functionally-grounded. They discuss when each type of study might be most appropriate, but do not elaborate on the individual design dimensions of the different types. \citet{hoffman_metrics_2018} provide four criteria for XAI evaluation: \emph{goodness, satisfaction, comprehension}, and \emph{performance} and focus on how these criteria can best be measured. 

\section{Literature Review}
\label{sec:literature_review}

\input{table.tex}

Many different buzzwords have been mentioned as goals for (X)AI in previous work: intelligibility, justifiability, or interpretability, to name just a few. However, it is not always immediately obvious how these goals can be achieved, and how success can be measured. In order to identify dimensions that have previously been evaluated and to distill guidelines for the evaluation of (X)AI, we conducted a literature review.

\subsection{Methodology}

\paragraph{Scope}
We have collected the proceedings from high-quality computer science journals and conferences. We include conference papers from ACM Computer-Human Interaction (CHI), ACM Intelligent User Interfaces (IUI), ACM Recommender Systems (RecSys) and IEEE VIS (VIS). Additionally, we include journal articles from IEEE Transactions on Visualization and Computer Graphics (TVCG) and Extended Abstracts published at CHI. Furthermore, we retrieved all publications from the International Conference on Machine Learning (ICML) and the ACM Conference on Fairness, Accountability, and Transparency (FAT*). For all venues, we considered the years 2015 to 2019 (2018 and 2019 for FAT*) to focus on recent developments.  

\paragraph{Paper Selection}
Once we had gathered the proceedings, we performed a keyword search for \emph{trust, interpretable, interpretability, explanation, explainability, transparency} and \emph{interactive machine learning} on the titles and abstracts of published works, retrieving an initial set of papers. We manually evaluated all potential papers of interest and excluded those that do not deal with some form of machine learning or artificial intelligence, or that do not perform user evaluation. For that reason, we excluded all papers from the FAT* and ICML from this review. Due to the large amount of papers, papers that did not include interactivity and those that covered relatively fewer dimensions were also excluded from our review. 

\paragraph{Coding}
We coded all papers in an iterative process and began with an initial set of eight randomly selected papers. After extracting all relevant dimensions and coding the initial paper set, we distilled and refined coding guidelines until an agreement between coders was reached. We then continued coding the remainder of papers with a single coder. Whenever we encountered new potential dimensions that had not been mentioned in papers previously coded, we conferred and decided whether to include them into adapted coding guidelines. During the coding process, significant differences between pure application papers and those with comparative study designs became apparent. We consequently decided to code application papers using separate guidelines (created using the same process) and present both types of papers separately in the following sections. This methodology allows for a more focused comparison of papers from a given paper type. 

\paragraph{Concept Definitions}
During paper coding, we did not attempt to resolve potential conflicts, ambiguities, or overlaps between concept definitions but coded them as presented by the authors. As a consequence, the results of our literature review present a ``union'' of the definitions for concepts like trustworthiness or interpretability. Refining these concepts and converging to a common vocabulary presents an opportunity for future work that will be elaborated on further in \autoref{sec:opportunities}.

\paragraph{Presentation}
The results of our literature review are summarized in \autoref{tab:survey_papers} (comparative study designs) and \autoref{tab:application_papers} (application papers). The tables highlight four different groups of design dimensions for structured experimental evaluation and sort them according to our trust building model introduced in \autoref{sec:bias_and_trust}. \textcolor{user}{\textbf{Personal}} contains both standard personal characteristics, as well as dimensions on experience. \textcolor{explanation}{\textbf{Explanation}}  and \textcolor{model}{\textbf{Model}} group dimensions of the respective elements in the XAI pipeline. \textcolor{te}{\textbf{Task \& Environment}} focuses on the implications of using a given (X)AI system in a specific environment.  Due to space constraints we do not include all dimensions that have been mentioned in literature. For example, \emph{controllability} and \emph{truthfulness} that were both mentioned only once were excluded. Instead, we focus on the most common dimensions and those that allow us to draw conclusions about the state of the field.

The tables also highlight the number of study participants, as well as the publication venue of the papers. For all dimensions, coloured boxes indicate whether they were \emph{study conditions}~\co, \emph{measured}~\m in a study, or \emph{fixed}~\f to an artificial value. Cases where dimensions varied as conditions where also measured, \cm is used. \OK and \NO ~indicate yes and no respectively, and are used to show whether explanations were available and whether the system was model agnostic from the point of view of the study participant. Possible values for task groups, as well as impact and criticality will be introduced in \autoref{sec:task_and_env}. Darker color (\quant{0} --- \quant{5}) indicates higher task complexity, impact or criticality, respectively. Similarly, we classify effort and user expertise that the system in an application paper was designed for as low~\dfquant{1}, medium~\dfquant{2} or high~\dfquant{3}. Some application papers claim that the presented systems are designed with a specific goal or property in mind, but do not evaluate their respective design decisions and are highlighted accordingly~\dff.

\subsection{Comparative Studies}
\label{sec:studies}

\autoref{tab:survey_papers} contains 21 studies from 19 publications.
While most publications are only present in the table once, \tablecite{yin_understanding_2019} provide three large-scale studies on the same subject and have thus been included three times. In the remainder of this section we briefly summarize the main findings of our literature review. 

\subsubsection{Summary of Findings}
17/21 studies include \emph{explanations} in their study design. Out of these 17, the availability or absence of explanations is a study condition. 
Most work only evaluates perceived \emph{trustworthiness of the machine learning model} (15/21) but not the \emph{trustworthiness} of the explanation (3/17 papers that include explanations). While this evaluation of trust in the model is essential, we note a distinct lack of evaluation of the trust in the model explanation. Such explanation evaluations are important in the light of trust-building and bias propagation, as modeled in \autoref{sec:bias_and_trust}. 

The inclusion of \emph{expertise} as a measured dimension appears to be a relatively recent development, with 7/8 studies having been published in 2019. Furthermore, the only reviewed study incorporating \emph{personality} traits was published in the same year. Knowledge about such user details should influence the \emph{information content}, the most utilized design dimension from the explanation group (9/17). This emphasizes the vast opportunities for presentation of explanations, including varying the level of detail or adding personalization.  

Only two studies investigate manipulating the fidelity of an explainer. 
Worryingly, \citet{eiband_impact_2019} find little difference in trust towards real or placebic explanations. 
Similarly, there seems to be little distinction between the reported understanding of participants, and real understanding that is proven through, for example, little tasks and quizzes.
This should inspire future research in that direction to avoid misleading users and miscalibrating their trust. 

\subsubsection{Discussion of Findings}
Many of the experiments evaluate trust in recommendations or social feeds. Those experiments mostly feature low impact and low criticality, making them appropriate for non-expert users. Nonetheless, evaluations of trust in higher-impact settings are needed, especially considering the typically higher criticality of expert-user applications (see \autoref{tab:application_papers}). Future studies are needed that draw from related work from psychology to adequately simulate scenarios that are more appropriate to real-world usage, instead of measuring individual variables in isolation. Better simulation of actual usage conditions and environments is likely to affect study results, especially when impact and criticality are high. 

As mentioned above, almost none of the studies evaluate the impact of explanation fidelity. While fidelity is arguably important for interpretability and trust building, its necessity varies depending on the target audience. In expert systems, explanations highlight the models decision-making processes and can uncover training issues or biases. Here, it is essential that all explanations be high-fidelity and follow the inner workings of a given model closely. For the explanation of social feeds or movie recommendations for casual users, however, explanation by example might be more intuitive and effective. As education about the model is not the primary goal, designers have more freedom when creating explanations. Nonetheless, ethical questions remain as wrong explanations can easily mislead users.

In a similar direction, background, age and gender of studies are not always reported, especially when they were conducted through online crowd-sourcing platforms. In addition to user expertise, these dimensions are likely to have a significant impact, though. In particular it is not clear how well studies conducted exclusively with participants from the US generalize to other user groups with large cultural differences. Such cultural differences are also likely to influence optimal information content of explanations. For example, previous work has found differences in the preference for personalized explanations depending on their cultural background~\cite{evans_psychology_2009}.  

\subsection{Application Papers}
\label{sec:application_papers}
\input{table_application.tex}

\subsubsection{Summary of Findings}
Application papers, almost by definition, describe the design of an XAI model and any accompanying evaluations. Only one paper (\citet{brooks_featureinsight:_2015}) makes the availability of the explanation a study condition. Most papers assess the explanations by applying them to a dataset as case studies or proof-of-concept demonstrations. Furthermore, most papers do not conduct any significant amount of testing on the explanation itself. For example, only one paper, (\citet{ming_understanding_2017}), discussed the fidelity of the explanation developed.

The systems presented in the reviewed application papers tend to support more complex tasks (such as model refinement or comparison) than the ones evaluated in the reviewed comparative studies. More than half of the papers also designed explanations for users with high ML expertise (18/35). This ties in to the low impact of most of these papers, since the explanations will necessarily only be relevant to machine learning experts rather than a broader demographic of users. Interestingly, accountability seems to be a more recent trend in XAI. Only one paper (\citet{cabrera_fairvis:_2019}) discussed accountability in terms of fairness and mitigating bias.

\subsubsection{Discussion of Findings}
Many of the reviewed application papers mention particular dimensions of (X)AI, such as trustworthiness, as design goals. However, these dimensions are rarely evaluated for in any user testing or case studies included in the papers. Without such evaluation, it would be harder to verify that the design criteria were indeed satisfied by the system created and that, for example, users indeed found the system to be trustworthy. As a consequence it would be difficult, going forward, to propose a set of guidelines for how (X)AI systems can be designed to meet certain criteria better. The same is true for user evaluations that are performed with a particularly low number of participants.

Finally, as mentioned above, many of the explanations presented in application papers are designed to be used by machine learning experts. In particular when the model being explained is used in the field of deep learning the machine learning experts using the explanations are often considered to be domain experts as well, regardless of the actual data domain. This suggests a potential area of research into designing explanations geared towards machine learning novices or individuals with different domain expertise.

\section{Design Dimensions for Experimental Evaluations}
\label{sec:dimensions}
In this section, we synthesize design dimensions for the structured experimental evaluation of explainable artificial intelligence from the literature review presented above. Where possible or necessary, we provide definitions. As many goals and properties of (X)AI have been defined in the literature but were not yet evaluated in the reviewed literature, we expand the dimensions with these definitions. All dimensions reuse the colors from \autoref{fig:teaser}. Higher opacity indicates dimensions that appear in \autoref{tab:survey_papers} or \autoref{tab:application_papers}, while lower opacity is used for all remaining dimensions. 

Previous work from \citet{doshi-velez_towards_2017} characterizes (X)AI along the dimensions of \emph{global} and \emph{local interpretability}, \emph{time limitation} and the nature of \emph{user experience}.~\cite{doshi-velez_towards_2017}. We have also identified those dimensions from our literature review and report them below. \citet{guidotti_survey_2018} have identified \emph{reliability, robustness, causality, scalability,} and \emph{generality} as desired dimensions for machine learning models~\cite{guidotti_survey_2018}. As these are high-level concepts, they have not yet been experimentally evaluated. They can, however, likely be approximated by the dimensions reported below.

\subsection{User Attributes}
\label{sec:user_attributes}

As our literature review emphasized the evaluation of trust in (X)AI, user attributes play an important role. This design dimension is characterized by the question: \textcolor{user}{\textbf{Who} was the (X)AI method \textit{designed for?}} Within the dimension, we distinguish between immutable personal characteristics and personal experience that is dependent on the circumstances. When designing experiments that modulate these dimensions, researchers can draw from extensive related work from psychology and the humanities on trust-building, explanation processes, and conversational explanations.

\paragraph{Personal Characteristics}
Personal characteristics are immutable. 

\begin{description}
    \bulletitem{Age}{user} The age group that the tool or system was designed for. We did not encounter work specifically targeting a certain age group. This provides opportunities for future research, for example in evaluating trust of teenagers in social media recommendations. 
    \bulletitem{Gender}{user} The gender that the system was designed for. Typically, we expect systems to be designed for fifty percent female users. Some studies report gender-specific trust measurements. \cite{kleinerman_providing_2018}
    \bulletitem{Background}{user} Cultural differences have a large influence on how we cooperate with peers, including machines, and how likely we are to follow or reject recommendations. Previous work from psychology found significant differences between some groups, but not others~\cite{evans_psychology_2009}. Opportunities for future work include verifying whether these findings transfer to trust in (X)AI.
    \bulletitem{Personality}{user} A dimension that is only mentioned in few studies and likely correlates with information captured by the \emph{background} dimension. Personal characteristics of interest include, among others, the propensity to trust, differences between trust in humans and machines, prejudice built from previous experience, confidence or self-esteem. 
    
\end{description}

\paragraph{Experience}

Objective assessment of experience is a challenging task, not only due to the Dunning-Kruger effect~\cite{dunning_chapter_2011} causing non-experts to be notoriously bad at rating their own experience. Instead, study designs should rely on asking questions about the number of years in a given field and testing participants' knowledge with questions. Participants can then be classified as \emph{novice, intermediate, proficient,} or \emph{expert} to make studies more easily comparable.

\begin{description}
    \bulletitem{Domain Expertise}{user} The visualization community often considers \emph{expert} domain knowledge from, for example, medicine, linguistics, or biology. However, ``casual users'' also have relevant domain knowledge, for example, in music or movies. Studies should investigate whether there are significant differences in trust between these two user groups and whether results from one are directly informative for designs targeting the respective other group.
    \bulletitem{Technical Expertise}{user} Technical experience includes general familiarity with computers or automation, as well as awareness of potential issues that may arise. Users that are more familiar with technology are generally expected to be more proficient at using (X)AI systems.
    \bulletitem{ML Expertise}{user} More specific than technical expertise, machine learning expertise is concerned with the familiarity with and understanding of the specific machine learning algorithms used. 
\end{description}

\subsection{Explanations}
\label{sec:explanations}

Depending on the user group, explanations might be necessary or not. If they are presented to users, care has to be taken to calibrate trust and avoid biases. This dimension is thus characterized by these questions:  \textcolor{explanation}{\textbf{Are} system decisions \textit{explained}? If so, \textbf{how}? } This group of dimensions can draw from a significant body of related work from social sciences. Not only do social sciences provide models of explanation, they also characterize expectations towards the explanation process~\cite{miller_explanation_2019}.

\begin{description}
    \bulletitem{Availability}{explanation} Many study designs include conditions without explanations as baselines. This is especially important when it is unclear whether explanations have an influence on a given variable or dimension. 
    \bulletitem{Information Content}{explanation} Once the general usefulness of explanations in a given scenario has been demonstrated, study designers have vast opportunities in varying the information content of explanations. Subdimensions include, among many others, the information density, personalization of explanations, the use of emotional or factual statements.
    \bulletitem{Trustworthiness}{explanation} The trustworthiness assigned to explanations by study participants. The trustworthiness of an explanation can be explicitly affected by manipulating the correctness of the explanation or, more subtly, the tone in which it is presented. 
    \bulletitem{Effectiveness}{explanation} Some studies measure the effectiveness of an explanation. This dimension is mostly used to capture the convincingness of an explanation to perform a given action, not in explaining some complex underlying theory.
    \bulletitem{Fidelity}{explanation} This dimension captures whether, and how well, an explanation actually explains the models' decision-making process (high fidelity), or just contains some information that is presented in the style of an explanation but does not correspond to the model in any way (low fidelity). High fidelity explanation methods are fundamental for effective (X)AI.
    \bulletitem{Strategy}{explanation} Three major reasoning strategies are known from the social sciences and used in (X)AI: inductive (example-based), deductive (theory-based), and abductive (inductive reasoning in the absence of all facts; iterative process once more knowledge becomes available) reasoning~\cite{elassady_structuring_2019}.
    \bulletitem{Transparency}{explanation!60}
    An explanation method is transparent when all its decision making processes can be observed and understood by users. While early work in (X)AI equated transparency with an explanation, later work found that transparency might be overwhelming~\cite{poursabzi-sangdeh_manipulating_2018}.
\end{description}

\subsection{Models}
\label{sec:models}

So far, dimensions have focused on users and model explanations. This group of dimensions characterizes models in detail and answers the question: 
\textcolor{model}{\textbf{Which} AI models are used in this process?}

\begin{description}
    \bulletitem{Agnostic}{model} Some systems are model-agnostic from the point of view of the study participant in the sense that the users do not know what model is powering the system if there is one at all. Designing a model-agnostic system or providing model details to the participant has implications for user awareness, primes them by setting expectations, and thus influences trust and biases. 
    \bulletitem{Observed Quality}{model} The quality of the model that users interact with, typically represented in terms of the accuracy that users could observe during the study on the actual data points used. 
    \bulletitem{Quality}{model} The actual quality of the model that users interact with. This quality is typically represented by the \emph{accuracy} measured on the held-out test data. Showing this number to study participants before or during the study sets their expectations, starting a new trust calibration process whenever observed quality and presented model quality differ. 
    \bulletitem{Transparency}{model} A model is transparent when all its decision-making processes can be observed and understood by users. While early work in (X)AI equated transparency with an explanation, later work found that transparency might be overwhelming~\cite{poursabzi-sangdeh_manipulating_2018}.
    \bulletitem{Correctness}{model} This dimension describes user-perceived correctness (as opposed to model quality) in terms of how well the system output aligns with users' expectations~\cite{rader_explanations_2018}.
    \bulletitem{Interpretability}{model} We define a system to be interpretable when users can understand why it behaves in a given way under given circumstances. In that sense, interpretability can be considered an inductive process, where users first create a mental model of the system and then verify whether the system is consistent with that mental model, making it interpretable. \citet{lipton_mythos_2018} has previously surveyed interpretability and suggests ``that interpretability is not a monolithic concept, but in fact reflects several distinct ideas.''
    \bulletitem{Accountatability}{model} Accountability ``measure[s] the extent to which participants think the system is fair and they can control the outputs the system produces.''~\cite{rader_explanations_2018}
    \bulletitem{Trustworthiness}{model} This high-level dimension is based on multiple other dimensions. A model can be considered trustworthy when it is correct (according to user beliefs) and interpretable. 
    \bulletitem{Simulatability}{model!60} A model is simulatable when users can successfully predict the model output for a given input. 
\end{description}

\begin{figure*}[h!]
  \includegraphics[width=\textwidth]{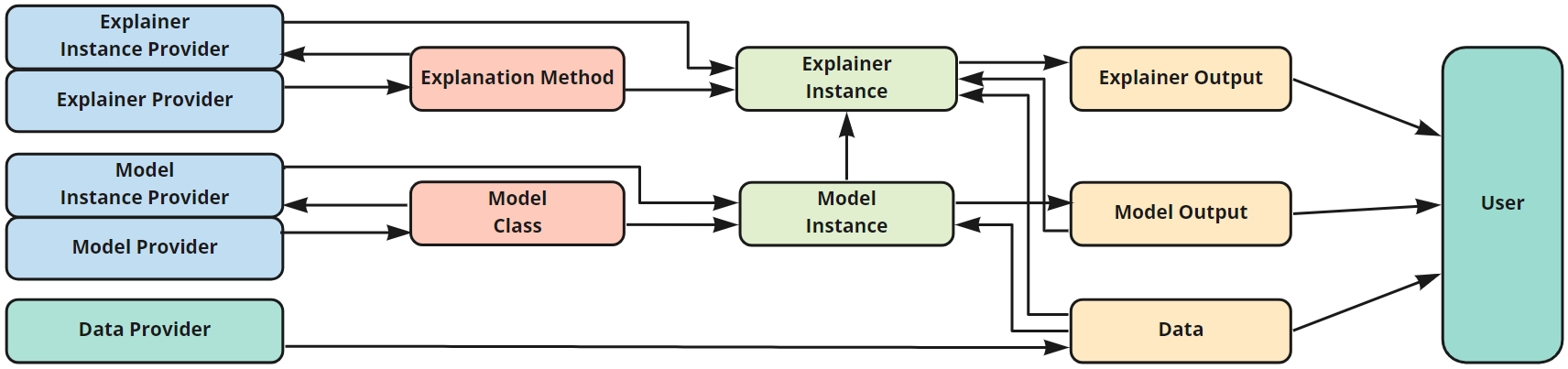}
  \caption{ Dependency Model for the XAI process. Bias propagates along the arrows, while trust is built based on the user's interaction with the data, model, and/or explainer outputs, respectively, following the dependency arrows in reverse.}
  \label{fig:model}
\end{figure*}

\subsection{Tasks and Environment}
\label{sec:task_and_env}

This group captures dimensions answering the question: \textcolor{te}{\textbf{How} and \textbf{where} are models and explanations used? } The dimensions characterize typical user tasks and the costs of (mis-)using an AI system. The cost is a combination of \emph{impact} and \emph{criticality}. We introduce some examples of different impact-criticality combinations after defining the dimensions.

\begin{description}
    \bulletitem{Task Group}{te} We identified seven general tasks with increasing difficulty: \textit{application} \quant{A}, \textit{understanding} \quant{U}, \textit{simulation} \quant{S}, \textit{diagnosis} \quant{D}, \textit{refinement} \quant{R}, \textit{justification} \quant{J}, \textit{comparison} \quant{C}. On the one hand, these groups simplify comparing system designs. On the other hand, it is apparent that users simply applying a machine learning model have different requirements towards explanations than those who have to diagnose problems, or justify decisions. Furthermore, previous work suggests that users are more likely to accept recommendations when working on complex tasks.~\cite{gino_effects_2007}
    \bulletitem{Impact}{te} The amount of people impacted by decisions supported through (X)AI systems likely has an influence on user behaviour: the more people are affected the less risk is acceptable. We define five categories for characterizing impact: \textit{none} \quant{0}, \textit{one} \quant{1}, \textit{some} \quant{2}, \textit{many} \quant{3}, and \textit{all} \quant{4}.
    \bulletitem{Criticality}{te} Criticality reports how severe the influence of an (X)AI system can be on those impacted. Possible values are \textit{none} \quant{0}, \textit{marginal} \quant{1}, \textit{significant} \quant{2}, \textit{troublesome} \quant{3}, \textit{livelihood} \quant{4}, \textit{extreme} \quant{5}.
    \bulletitem{Effort}{te} Independent of the impact and criticality, human actions motivated by (X)AI systems require a certain amount of effort when executed. We classify this effort as \emph{low}~\quant{1}, \emph{medium}~\quant{2} or \emph{high}~\quant{3}. Related work from psychology shows significant differences in willingness to act depending on the effort required~\cite{langer_mindlessness_1978}.
\end{description}

A few years ago, a bug in the voice assistant Alexa caused smart speakers in many people's homes to play laughs randomly. We classify this incident as high-impact, low-criticality as it affected many people but caused no harm. In contrast, the failure of the autopilot of a car or plane would be high-criticality, and low- or high-impact, respectively. 

\subsection{Study Design} 
\label{sec:study_design}
In addition to model- and user-specific dimensions, evaluation studies are highly dependent on the study design and setup. This section thus groups dimensions answering the question: \textcolor{gray}{\textbf{How} was the study designed?}

\begin{description}
    \bulletitem{Participant Age}{gray} The age of the study participants. This age can differ from the age group that a system was designed for.
    \bulletitem{Exerted Effort}{gray!60} The effort that was needed to complete the study. In order to produce reliable results that can be generalized, the gap between the effort needed in actual use and the effort exerted under study conditions should be minimized. 
    \bulletitem{Study Constellation}{gray!60} Under study constellation, we summarize all variables like the number of participants completing the study in parallel, whether participants were intentionally disturbed or distracted to create the desired effect, or how much help was available, for example, in pair analytics sessions. This dimension offers great potential for creating realistic study settings that replicate real-world usage conditions.  
    \bulletitem{Time}{gray!60} Many (X)AI systems are used under time pressure in day-to-day operations. Consequently, evaluation studies need to be run under realistic time limits to create a comparable environment. 
    \bulletitem{Trust Measure}{gray!60} There are different ways to measure trust. \citet{poursabzi-sangdeh_manipulating_2018} use simulation and weight of advice for prediction tasks, but different systems might require different trust measures.
\end{description}

\section{Structuring the Design Space} 
\label{sec:bias_and_trust}

To generate a holistic view on the evaluation of XAI we strive to bring the design dimensions into context, structuring  the design space of XAI studies and defining their scope and influences.
In this section, we present a dependency model for XAI processes. This model describes the different stages and stakeholders of XAI. Each of the dimensions detailed in \autoref{sec:dimensions} has a different impact on the model's components. In particular, our contextual model can be utilized to describe biases that might arise within XAI processes, as well as bias propagation through the dependencies. 
In addition, we postulate that trust building occurs through the user's interaction with the data, AI model, and explanations. Hence, trust building follows the dependency arrows of our model in reverse order. This section emphasizes that the aim of XAI studies should be to achieve a broad coverage, while not increasing the complexity of the modeling or loosing too much detail. In the following, we describe the dependency model in more detail, and discuss the processes of bias propagation and trust building in (X)AI.

\subsection{Dependency Model}
The proposed, conceptual dependency model in \autoref{fig:model} covers the stakeholders in XAI systems and the building blocks they provide. This model highlights \textit{dependencies} and is not designed to model possible \textit{interactions} or \textit{iterative feedback loops}. In the following, we describe our dependency model as depicted in \autoref{fig:model}. We simply refer to AI/ML models as ``\textit{model}'' in the remainder of this section. 

The \emph{\textbf{model provider}} is a person or entity creating a novel \emph{\textbf{model class}}. Such a model class is subject to the design dimensions presented in \autoref{sec:models}. A concrete \emph{\textbf{model instance}} is needed for the model to be practically used. Such an instance is created by the \emph{\textbf{model instance provider}}. In addition to a model implementation (omitted in the model to avoid unnecessary complexity unrelated to bias and trust propagation) the instance provider typically requires some \emph{\textbf{training data}}. This data comes from the \emph{\textbf{data provider}}. Note that all stakeholders in the model might be the same entity, or all be distinct. Once the model is trained, it can produce some \emph{\textbf{model output}}. Together with the training data and the model instance, this output forms the potential inputs for a model \emph{\textbf{explainer instance}}. Which inputs are actually used depends on the type of the explainer~\cite{spinner_explAIner_2019}. Analogous to models, explainers are particularly influenced by dimensions from \autoref{sec:explanations}, and have an associated \emph{\textbf{explainer provider}}, \emph{\textbf{explainer method}} and an \emph{\textbf{explainer instance provider}}. Depending on the system design, \emph{\textbf{explainer output}}, model output and training data, data might be available to the user, who is characterized by dimensions from \autoref{sec:user_attributes}. 

\subsection{Bias Propagation}
Despite systems and explanations being designed as deliberately as possible, they are still subject to external factors like where or by whom a machine learning model has been trained or deployed. The dependency model contains many stakeholders with potentially diverging goals and interests: a data provider might discriminate against foreigners for political reasons, a model instance provider against minorities, and a particular explainer might only be useful to experts. Whether it is willingly or unwillingly,  such biases might hamper the trustworthiness of the complete (X)AI pipeline.  

As these potential biases \textit{propagate} through the XAI process, we use the dependency model to describe their influences. 
For example, it is impossible to obtain a fair and unbiased, high-quality model if the training data had a racial or gender bias. Increasing the transparency of a system, for example through explanations, can help to reveal such biases; however, it does not reduce them. On the contrary, explanations might miscalibrate user trust in a system and lessen bias awareness. This is even true for high-fidelity explanations that correctly represent the model's decision-making process: if the model itself is insufficient, any local explanation can itself be correct, while still misleading users and not revealing model shortcomings. We, therefore, argue for acknowledging sources of potential biases and their effects on other stages and stakeholders in the XAI process using the dependency model.  Mitigating such biases in-place and limiting their propagation through the model can reduce their harmful effects, as mentioned above. 

In addition to those general biases that are present for all users, the (X)AI process is subject to user-specific biases. Those biases are based on users' previous experiences and their knowledge. In our model, biases can be added to the process at any block in \autoref{fig:model}. There, they will \textit{increase}  the existing biases and propagate along the depicted dependency arrows. The effects of bias propagation in (X)AI are an interesting area of future research and studies.

\subsection{Trust Building}
Humans usually do not build trust in abstract concepts, but concrete outputs that they see and can potentially interact with. Consequently, users first recognize that a model seems to work well and that the explanations seem to make sense. Once they have built trust in the explanations and model outputs, they start building trust in the model, before eventually trusting the model instance provider.
As users propagate back through the XAI process, 
their personal biases apply, influencing trust building positively or negatively. 

For example, a particular user might have experienced unreliability using deep learning classifiers resulting in a personal conviction that such models do not work. This person might also have had a good experience using models provided by a particular tech company, resulting in a positive personal bias towards the model provider. If in our case, this user provides their own data that they are familiar with, they can judge the trustworthiness of a model output given their expectations about the data. They can probably also judge the fidelity of an explainer based on its output. Using our dependency model, we can follow the trust-building process,  taking into account positive and negative amplifications that are reinforced through biases. If the explanations provided by the XAI method are matching the user's expectation, they might foster trust in the explainer output and start building trust in the explainer instance. This effect might be reinforced by other factors, such as a positive experience using the same explainer instance on a different data point. Increasing the trust in the explainer instance might lead to the user starting to trust the explainer instance provider and/or the explanation method itself.

To answer the question of whether or not we should trust (X)AI, we have to take the cross-relational effects of the dependency model into account. In particular, when designing evaluations for (X)AI systems, we have to consider the coverage of the model. In our literature review, we broadly observed a disconnect between the different research communities; with HCI focusing more on trust-building in the presentation of  X(AI), and the AI/ML community, generally concerned with the correctness of the models as a means to increase trust. We argue that a broad coverage of the XAI process is necessary, as observed in some application papers (within their focused scope). Moreover, we postulate that studies concerned with one part of the dependency model should abstain from partly including descriptions of other parts of the model without considering possible dependencies and cross-effects. Coming back to our example from above this means that a study participant should not be informed about the model providers (the tech company) if the study design is not set up to appropriately capture potential biases, their dependencies, and the resulting effects on trust.

\section{Discussion and Implications}
\label{sec:opportunities}

The model presented in the previous section has direct implications for study design: any components of the model that are mentioned in the study prototype, questionaire, or some meta-information must be taken into account as potential sources of bias, distorting results. At the same time, they highlight the vast opportunities for future work, conducting studies that include or exclude those particular areas of the model. In the following, we will present further opportunities, as well as limitations of our work. 

\subsection{Opportunities}

\begin{description}
\item[(1)~Strive Towards Better Coverage of the XAI Process] The dependency model highlights stakeholders in the different stages of the (X)AI process and their dependencies. Currently, the different communities tend to focus on different stages of the process when conducting evaluations: the machine learning community does not typically involve users in evaluations, and the HCI community tends to focus on the presentation of model outputs and explanations. Better coverage of the model in the form of studies spanning multiple stages is required to bring the field forward. While application studies (mostly from VIS) attempt to bridge this gap by providing rich model interactions, they are problem-specific and often only gather qualitative feedback. Consequently, there is great potential for collaboration between these communities to provide end-to-end testing of the (X)AI process, explaining the inner workings of machine learning models.

\item[(2)~Bridge the Gap to other Communities]
In addition to better connecting the different communities from computer science, evaluation of (X)AI can profit from collaborations with the social sciences, and previous sections have already occasionally alluded to particular related work from psychology. Significant bodies of work have investigated trust in inter-human relations, and made generalizations towards human-robot~\cite{hancock_meta-analysis_2011} and human-automation collaboration~\cite{schaefer_meta-analysis_2016}. The respective experiments should be repeated to verify that their findings still apply for human-AI collaboration. Additionally, collaboration with psychologist is needed to create study scenarios that more closely resemble real-world usage conditions, rather than mostly relying on online crowd-sourced studies. This is especially important for evaluating use-case specific applications tackling high-impact and high-criticality issues.

\item[(3)~Apply a Clearly Defined Terminology]
We observed a tendency, among some papers that we studied, of stating some systems were designed with specific goals like interpretability in mind, but not evaluating whether the said properties were achieved. More work like the structuring review by \citet{lipton_mythos_2018} is needed to refine and merge the concepts that have already been proposed.  This allows related fields to converge on common terminologies that are well-aligned with each community's identified goals.  Then, instead of defining more high-level goals for (X)AI, researchers would consider how the proposed dimensions can be measured effectively, and what approximations and proxies might be necessary.

\item[(4)~Acknowledge Biases and Propagation of Trust]
Researchers evaluating (X)AI should acknowledge the inherent biases in human trust-building and draw from related work in the social sciences. This should, in particular, influence the presentation of explanations. Results from psychology show that ``sets of source factors (expertise, liking, trust, and similarity) and message factors (politeness, response efficacy, feasibility, absence of limitations, and confirmation)''~\cite{feng_influences_2010} each influence how humans deal with advice. Especially \emph{liking} and \emph{trust} are likely to vary from individual to individual based on existing prejudices and biases. 

\item[(5)~Trust in Explanations vs. Trust in Models]
Typical XAI systems aim to increase trust by providing explanations. Success is then often measured by evaluating the trust users have in the system. However, if the explanations are not transparent to the user and cannot be verified, they cannot ease doubts about the correctness of the underlying model. Instead, they simply shift the problem to a different stage of the XAI process by explaining a black box with a black box. Consequently, the trust in explanations and the correct calibration of trust in them should take a more central role in XAI research. 

\item[(6)~Consider Explainer Fidelity]
Model complexity, especially of deep neural networks, is quickly increasing thanks to the availability of relatively cheap computing resources. The more difficult it becomes for humans to interpret these complex models, the more difficult it becomes to generate intuitive explanations. At the same time, some work from psychology on placebic explanations has already been successfully replicated in the context of XAI~\cite{langer_mindlessness_1978, eiband_impact_2019}, and it has been suggested that humans are eager to believe explanations they are provided with~\cite{hohman_gamut:_2019}. Consequently, a cornerstone of XAI research should be ensuring that explainers have high fidelity. Otherwise, the field risks producing explanations that ``sound good'' but are misleading users and exploiting miscalibrated trust. 

\item[(7)~Incorporate Motivation for Use]
There can be various reasons for users to be interacting with (X)AI. These reasons can stem from intrinsic or extrinsic motivation, be spontaneous or persistent over time. All these factors influence the perceived impact and criticality of a given task, and consequently, the effort that users are willing to exert. Not only is it important to motivate system users, for example, using gameful design elements~\cite{sevastjanova_gameful_2019}, but also to create study constellations that reflect those motivations. 
\end{description}

\subsection{Limitations}

To set the scope of our work, in this section, we acknowledge the limitations of our literature review and discuss potential alternatives. First, our review is based on a keyword search on the title and abstract of impactful venues. Future work could extend this review into a survey by expanding on the methodology and including more related work through the inclusion of forward and backward references.  
As previously elaborated by \citet{lipton_mythos_2018}, not all concepts in XAI are clearly defined. We do not attempt a disambiguation of terms used. Instead, our coding is directly based on the concepts mentioned in the respective reviewed work. Further, we only collect and review relevant papers and do not provide a complex meta-evaluation taking reported significant differences and effect sizes into account. Instead, we focus on the reported dimensions. The dependency framework presented in \autoref{sec:bias_and_trust} is preliminary and focuses on those entities that are important for trust building and bias propagation. Future work should provide extensions covering model implementations (a common source of errors from our experience) and afforded interaction possibilities. 

\subsection{Future Work}
Our literature review revealed the independent goals of various sub-domains in the computer science community. Visual analytics, mostly working with expert systems, is concerned with explanations teaching users the inner workings of models and educating them in machine learning, while recommender systems are often tailored to be convincing. This work presents general design dimensions that are applicable to various domains. Tailoring those dimensions to the specific goals of the domains facilitates converging towards a common, shared vocabulary as introduced above.

As \autoref{tab:survey_papers} revealed, the majority of comparative studies assess the trustworthiness of models. Most of the reviewed works asked study participants to rate the trustworthiness on Likert-scale questionnaires. Instead of adding this step of indirection, \citet{yin_understanding_2019} utilize simulatability and weight of advice as proxies for trust. Avoiding post-usage questionnaires and relying on implicit trust measures lowers the risk of users' preconceptions and biases influencing their trust-responses. Future work should investigate which proxies for trust are best suited for which tasks.

\section{Conclusion}

We have presented a literature review of the past five years of explainable artificial intelligence in the visualization and human-computer-interaction communities. From our review, we have distilled design dimensions for the user-centered evaluation of (X)AI methods and systems. Comparing those design dimensions and goals typically mentioned for (X)AI, we identified research gaps and opportunities for future work. So far, many design dimensions have barely been utilized in evaluations. This is especially true for abstract concepts like interpretability or accountability. We have also presented a dependency model highlighting the different stages of the (X)AI process and showing how bias and trust propagate in (X)AI systems. Together with the design dimensions, this model guides future evaluations of (X)AI systems. 

\bibliography{sample-base}

\end{document}

%% file: table.tex
\begin{table*}[htb]
    \centering
    \definecolor{blueish}{HTML}{F4F7F8}
    \rowcolors{2}{white}{blueish}
    \setlength{\tabcolsep}{1.5pt}
    \renewcommand\arraystretch{1.20}
    \begin{tabular}{r | *{4}{c} | *{6}{c} | *{8}{c} | *{4}{c} | *{2}{c} |}
        & \multicolumn{4}{|c|}{\textcolor{user}{\textbf{User}}} & \multicolumn{6}{c|}{\textcolor{explanation}{\textbf{Explanation}}} & \multicolumn{8}{c|}{\textcolor{model}{\textbf{Model}}} & \multicolumn{4}{c|}{\textcolor{te}{\textbf{T \& E}}} & \multicolumn{2}{c|}{\textbf{Misc.}} \\
        \cmidrule{2-25}
        Publication & \rot{Domain Expertise}  & \rot{ML/AI Expertise} & \rot{Tech. Expertise} & \rot{Personality} & 
        \rot{Available?} & \rot{Information Content} & \rot{Trustworthiness} & \rot{Effectiveness} & \rot{Fidelity} & \rot{Strategy} & 
        \rot{Agnostic?} & \rot{Quality / Accurracy} & \rot{Observed Quality} &  \rot{Transparency} & \rot{Trustworthiness} & \rot{Correctness} & \rot{Interpretability} & \rot{Accountability} &
        \rot{Task Group}  & \rot{Impact} & \rot{Criticality} & \rot{Effort} &
        \rot{\# Participants} & \rot{Venue} \\
        
        \midrule
        \tablecite{cai_effects_2019} &
        & & & &
        \OK & \co & & & & &
        \OK & & & & & & & &
        \NO & \quant{0} & \quant{1} & \quant{1} & 
        1070 & IUI\\
        
        \tablecite{cheng_explaining_2019} &
        & \m & \m & &
        \co & & & & &  &
        \NO & & & & \m & & & &
        \quant{A} & \quant{3} & \quant{3} & \m &
        202 & CHI\\
        
        \tablecite{dodge_explaining_2019} &
        & & & &
        \OK & \co & & & & \co & 
        \OK & & & & \m &  & & &
        \quant{A} & \quant{1} & \quant{1} & \quant{1} & 
        160 & IUI\\
        
        \tablecite{dominguez_effect_2019} &
        & & & &
        \OK & \co & & & & & 
        \OK & \m & & & \m &  & & &
        \quant{A} & \quant{1} & \quant{1} & \m & 
        121 & IUI\\
        
        \tablecite{eiband_impact_2019} &
        & & & &
        \OK & \cm & \m & \m & \co & & 
        \OK & & & & \m &  & & &
        \quant{A} & \quant{1} & \quant{1} & \quant{1} & 
        30 & CHI EA\\
        
        \tablecite{kouki_personalized_2019} &
        \m & & & \m &
        \OK & \co & & \m & & & 
        \OK & \m & & \m & \m &  & & &
        \NO & \quant{1} & \quant{1} & \quant{1} & 
        198 & IUI\\
        
        \tablecite{millecamp_explain_2019} &
        \m & & \m & &
        \co & & & & & & 
        \OK & \m & & & \m & & \m & &
        \quant{A} & \quant{1} & \quant{1} & \quant{1} & 
        71 & IUI \\
        
        \tablecite{richter_effects_2019} &
        & & & &
        \OK & & & \m & & & 
        \OK & & & & &  & & &
        \quant{A} & \co & \co & & 
        65 & IUI\\
        
        \tablecite{schaffer_i_2019} &
         & & \m & &
        \co & & \co & & & & 
        \OK & & & & &  & & &
        \NO & \quant{1} & \quant{1} & \quant{1} & 
        551 & IUI\\
        
        \tablecite{springer_progressive_2019} &
      & & & &
        \co & & & & & & 
        \OK & & & \m & \m &  & & &
        \NO & \quant{1} & \quant{1} & \quant{1} & 
        74 & IUI\\
        
        \tablecite{yin_understanding_2019} &
      & & & &
        \NO & & & & & & 
        \NO & \co & \f & & \m &  & & &
        \quant{S} & \quant{1} & \co & \quant{1} & 
        1994 & CHI\\
        
        \tablecite{yin_understanding_2019} &
      & & & &
        \NO & & & & & & 
        \NO & \co & \f & & \m &  & & &
        \quant{S} & \quant{1} & \co & \quant{1} & 
        757 & CHI\\
        
        \tablecite{yin_understanding_2019} &
      & & & &
        \NO & & & & & & 
        \NO & \co & \f & & \m &  & & &
        \quant{S} & \quant{1} & \co & \quant{1} & 
        1042 & CHI\\
        
        \tablecite{zhou_effects_2019} &
      & & & &
        \co & & & & \co & & 
        \NO & & & & \m &  & & &
        \quant{D} & \quant{1} & \quant{1} & & 
        22 & CHI EA\\
        
        \tablecite{bigras_working_2018} &
      & & & & 
        \OK & \co & & & & & 
        \OK & & & & \m &  & & &
        \quant{A} & \quant{3} & \quant{2} & \co & 
        20 & CHI EA\\
        
        \tablecite{kleinerman_providing_2018} &
      & & & &
        \OK & \co & & & & & 
        \OK & & \m & \m & \m &  & & &
        \NO & \quant{1} & \quant{1} & \quant{1} & 
        59 & RecSys\\
        
        \tablecite{rader_explanations_2018} &
      \m & & & &
        \OK & & & & & & 
        \NO & & & & & \m & \m & \m &
        \quant{A} & \quant{1} & \quant{1} & & 
        681 & CHI\\
        
        \tablecite{yu_user_2017} &
      & & & &
        \NO & & & & & & 
        \NO & \m & & & \m & \co & & &
        \quant{A} & \quant{2} & \quant{2} & \quant{1} & 
        21 & IUI\\
        
        \tablecite{chang_crowd-based_2016} &
      & & & &
        \OK & \cm & \m & & & & 
        \OK & & & & &  & & &
        \NO & \quant{1} & \quant{1} & \quant{1} & 
        220 & RecSys\\
        
        \tablecite{kizilcec_how_2016} &
      & & & &
        \OK & & & & & & 
        \NO & & & \co & \m &  & & &
        \NO & & & & 
        103 & CHI\\
        
        \tablecite{musto_explod:_2016} &
      & & & &
        \OK & \co & \m & & & & 
        \NO & & & & &  & & &
        \NO & \quant{1} & \quant{1} & \quant{1} & 
        308 & RecSys\\
        \bottomrule
    \end{tabular}
    \caption[]{Synthesis of the most important dimensions mentioned in previous work on (explainable) artificial intelligence. Little squares indicate that a variable was artificially manipulated to a \emph{fixed}~\f value, \emph{measured}~\m, constituted a ~\emph{condition}~\co, or a combination thereof \cm. }
    \label{tab:survey_papers}
\end{table*}

%% file: table_application.tex
\begin{table*}[tbp]
    \centering
    \definecolor{blueish}{HTML}{F4F7F8}
    \rowcolors{2}{white}{blueish}
    \setlength{\tabcolsep}{1.5pt}
    \renewcommand\arraystretch{1.20}
    \begin{tabular}{r | *{4}{c} | *{7}{c} | *{8}{c} | *{4}{c} | *{2}{c} |}
        & \multicolumn{4}{|c|}{\textcolor{user}{\textbf{User}}} & \multicolumn{7}{c|}{\textcolor{explanation}{\textbf{Explanation}}} & \multicolumn{8}{c|}{\textcolor{model}{\textbf{Model}}} & \multicolumn{4}{c|}{\textcolor{te}{\textbf{T \& E}}} & \multicolumn{2}{c|}{\textbf{Misc.}} \\
        \cmidrule{2-26}
        Publication & \rot{Domain Expertise}  & \rot{ML/AI Expertise} & \rot{Tech. Expertise} & \rot{Personality} & 
        \rot{Available?} & \rot{Information Content} & \rot{Trustworthiness} & \rot{Effectiveness} & \rot{Fidelity} & \rot{Strategy} & \rot{Iterative} & \rot{Agnostic?} & \rot{Quality / Accurracy} & \rot{Observed Quality} &  \rot{Transparency} & \rot{Trustworthiness} & \rot{Correctness} & \rot{Interpretability} & \rot{Accountability} &
        \rot{Task Group}  & \rot{Impact} & \rot{Criticality} & \rot{Effort} &
        \rot{\# Participants} & \rot{Venue} \\
        
        \midrule
        \tablecite{brooks_featureinsight:_2015} &
        & \dfquant{2} & & &
        \m & & & & & & &
        \OK & \m & \dff & & & & \dff & &
        \quant{R} & \quant{2} & \quant{1} & \m & 
        & VIS\\
        
        \tablecite{cabrera_fairvis:_2019} &
        \dfquant{3} & \dfquant{3} & \dfquant{3} & &
        \OK & & & & & & &
        \OK & \dff & & & & & & \dff &
        \quant{D} & \quant{1} & \quant{2} & & 
        & VIS\\
        
        \tablecite{cavallo_clustrophile_2019} &
        \co & \co & \co & &
        \OK & & & & & & \OK &
        \OK & & & & & \m & \m & &
        \quant{U} & \quant{2} & \quant{1} & & 
        12 & VIS\\
        
        \tablecite{el-assady_progressive_2018} &
        \co & \co & \co & &
        \OK & & & & & & \OK &
        \OK & \m & \dff & & \m & & \dff & &
        \quant{R} & \quant{2} & \quant{1} & \quant{3} & 
        6 & VIS\\
        
        \tablecite{el-assady_visual_2019} &
        \co & \co & \co & &
        \OK & & & & & & \OK &
        \OK & \m & \dff & \dff & & \m & & &
        \quant{C} & \quant{2} & \quant{1} & \quant{1} & 
        6 & VIS\\
        
        \tablecite{el-assady_semantic_2019} &
        \co & \co & \co & &
        \OK & & & & & & \OK &
        \OK & \m & \dff & & & \m & & &
        \quant{R} & \quant{2} & \quant{1} & \quant{1} & 
        6 & VIS\\
        
        \tablecite{hohman_summit:_2019} &
        & \dfquant{3} & & &
        \OK & & & & & & &
        \NO & & & & & & \dff & &
        \quant{U} & \quant{2} & \quant{1} & & 
        & VIS\\
        
        \tablecite{kahng_activis:_2018} &
        \dfquant{3} & \dfquant{3} & \dfquant{3} & &
        \OK & & & & & & &
        \NO & & & & & & \dff & &
        \quant{R} & \quant{2} & \quant{1} & & 
        & VIS\\
        
        \tablecite{kahng_gan_2019} &
        & \dfquant{1} & & &
        \OK & & & & & & \OK &
        \NO & & & & & & \dff & &
        \quant{S} & \quant{3} & \quant{2} & \quant{1} & 
        & VIS\\
        
        \tablecite{krause_workflow_2017} &
        \dfquant{3} & \dfquant{3} & & &
        \OK & & & & & & &
        \NO & & \dff & \dff & & & & &
        \quant{R} & \quant{2} & \quant{1} & \quant{1} & 
        & VIS\\
        
        \tablecite{kumpf_visualizing_2018} &
        \dfquant{3} & & & &
        \OK & & & & & & \OK &
        \OK & & \dff & & & & & &
        \quant{S} & \quant{3} & \quant{2} & \quant{2} & 
        & VIS\\
        
        \tablecite{kwon_clustervision:_2018} &
        \dfquant{3} & \dfquant{3} & & &
        \OK & & & & & & \OK &
        \OK & & & & & & & &
        \quant{A} & \quant{2} & \quant{1} & & 
        & VIS\\
        
        \tablecite{kwon_retainvis:_2019} &
        \dfquant{3} & & \dfquant{3} & &
        \OK & & & & & & \OK &
        \OK & \m & & \dff & & & \dff & &
        \quant{S} & \quant{3} & \quant{4} & & 
        & VIS\\
        
        \tablecite{lin_rclens:_2018} &
        & \dfquant{3} & & &
        \OK & & & & & & &
        \NO & \m & & & & & & &
        \quant{D} & \quant{2} & \quant{1} & \quant{2} & 
        & VIS\\
        
        \tablecite{liu_towards_2017} &
        & \dfquant{3} & & &
        \OK & & & & & & \OK &
        \NO & \m & & & & & & &
        \quant{R} & \quant{2} & \quant{1} & & 
        & VIS\\
        
        \tablecite{liu_analyzing_2018} &
        & \dfquant{3} & & &
        \OK & \dff & & & & & &
        \NO & & & & & & & &
        \quant{D} & \quant{2} & \quant{1} & & 
        & VIS\\
        
        \tablecite{liu_understanding_2018} &
        \quant{3} & & & &
        \OK & & & & & & \OK &
        \NO & & & & \m & & & &
        \quant{A} & \quant{2} & \quant{4} & & 
        14 & VIS\\
        
        \tablecite{liu_visual_2018} &
        & \dfquant{3} & & &
        \OK & & & & & & \OK &
        \NO & \m & & & & & & &
        \quant{D} & \quant{2} & \quant{1} & & 
        & VIS\\
        
        \tablecite{liu_visual_2018-2} &
        & \dfquant{3} & & &
        \OK & & & \m & \dff & & &
        \NO & & & & & & \dff & &
        \quant{D} & \quant{2} & \quant{1} & & 
        & VIS\\
        
        \tablecite{liu_analyzing_2018-1} &
        & \dfquant{3} & & &
        \OK & & & & & & &
        \NO & \f & & & & & & &
        \quant{D} & \quant{2} & \quant{1} & & 
        & VIS\\
        
        \tablecite{ma_explaining_2019} &
        & \dfquant{3} & & &
        \OK & & & & & & &
        \NO & \m & & & & & & &
        \quant{D} & \quant{2} & \quant{1} & \quant{3} & 
        & VIS\\
        
        \tablecite{ming_understanding_2017} &
        & \dfquant{3} & & &
        \OK & & & \m & & & &
        \NO & \f & & & \dff & & \dff & &
        \quant{D} & \quant{2} & \quant{1} & \quant{1} &
        & VIS\\
        
        \tablecite{ming_rulematrix:_2019} &
        \dfquant{3} & & & &
        \OK & & & & \dff & & &
        \NO & \m & & \dff & & & \dff & &
        \quant{R} & \quant{2} & \quant{1} & & 
        9 & VIS\\
        
        \tablecite{muhlbacher_treepod:_2018} &
        \dfquant{3} & & & &
        \OK & & & & & & \OK &
        \OK & \m & \dff & & & & \dff & &
        \quant{R} & \quant{2} & \quant{1} & & 
        & VIS\\
        
        \tablecite{pezzotti_approximated_2017} &
        & \dfquant{3} & & &
        \OK & & & & & & \OK &
        \NO & \dff & & & & & \dff & &
        \quant{D} & \quant{2} & \quant{1} & & 
        & VIS\\
        
        \tablecite{ren_squares:_2017} &
        & \co & \dfquant{3} & &
        \OK & & & & & & &
        \NO & \m & \dff & & & & & &
        \quant{C} & \quant{2} & \quant{1} & \quant{1} & 
        24 & VIS\\
        
        \tablecite{sacha_somflow:_2018} &
        & \dfquant{3} & & &
        \OK & & & & & & &
        \OK & & & & & & & &
        \quant{A} & \quant{2} & \quant{1} & & 
        & VIS\\
        
        \tablecite{spinner_explAIner_2019} &
        & \co & & &
        \OK & & \m & & & & \OK &
        \OK & & & & & & \dff & &
        \quant{R} & \quant{3} & \quant{1} & \quant{3} & 
        9 & VIS\\
        
        \tablecite{stahnke_probing_2016} &
        & \co & & &
        \OK & & \dff & \m & & & &
        \NO & & \dff & & & & & &
        \quant{U} & \quant{2} & \quant{1} & \quant{3} & 
        & VIS\\
        
        \tablecite{stoffel_feature-based_2015} &
        & \dfquant{3} & & &
        \OK & & & & & & \OK &
        \OK & & & & & & & &
        \quant{R} & \quant{2} & \quant{1} & & 
        & VIS\\
        
        \tablecite{strobelt_lstmvis:_2018} &
        & \dfquant{3} & & &
        \OK & & & & & & &
        \NO & & & & & & \dff & &
        \quant{D} & \quant{3} & \quant{1} & & 
        & VIS\\
        
        \tablecite{strobelt_seq2seq-vis:_2019} &
        & \dfquant{3} & & &
        \OK & & & & & & \OK &
        \OK & & & & & & & &
        \quant{R} & \quant{2} & \quant{1} & & 
        & VIS\\
        
        \tablecite{wang_dqnviz:_2019} &
        & \dfquant{3} & & &
        \OK & & & & & & \OK &
        \OK & \m & & & & & & &
        \quant{R} & \quant{2} & \quant{1} & & 
        & VIS\\
        
        \tablecite{zhang_manifold:_2019} &
        & \dfquant{3} & & &
        \OK & & & & & & \OK &
        \NO & \m & & \dff & & & \dff & &
        \quant{R} & \quant{2} & \quant{1} & & 
        & VIS\\
        
        \tablecite{zhao_iforest:_2019} &
        & \dfquant{2} & & &
        \OK & & & \m & & & &
        \NO & & & \dff & & & \dff & &
        \quant{U} & \quant{3} & \quant{1} & \quant{1} & 
        & VIS\\
        \bottomrule
    \end{tabular}
    \caption[]{The most important dimensions mentioned in previous application work on (explainable) artificial intelligence. Little orange squares indicate that a system was designed with a goal or property of (X)AI in mind, and that the property was evaluated~\f or not evaluated~\dff. \emph{Measured}~\m variables and experimental \emph{conditions}~\co are also shown.}
    \label{tab:application_papers}
\end{table*}